\documentclass[aps,prb,twocolumn,groupedaddress]{revtex4-2} 

\usepackage[]{graphicx}
\usepackage[dvipsnames]{xcolor}
\usepackage[colorlinks=true,citecolor=NavyBlue,linkcolor=RubineRed,urlcolor=Cerulean,pdfencoding=auto]{hyperref}
\usepackage[]{amsmath}
\usepackage[]{amssymb}

\def\bra#1{\mathinner{\langle{#1}|}}
\def\ket#1{\mathinner{|{#1}\rangle}}

\def\abs#1{\mathinner{|{#1}|}}

\def\comm#1#2{\mathinner{[{#1},{#2}]}}

\begin{document}

\title{Hyperfine Structure of Transition Metal Defects in SiC}

\author{Benedikt Tissot}
\email[]{benedikt.tissot@uni-konstanz.de}

\author{Guido Burkard}
\email[]{guido.burkard@uni-konstanz.de}

\affiliation{Department of Physics, University of Konstanz, D-78457 Konstanz, Germany}


\begin{abstract}
 Transition metal (TM) defects in silicon carbide (SiC) are a promising platform in quantum technology, especially because some TM defects emit in the telecom band.
  We develop a theory for the interaction of an active electron in the \(D\)-shell of a TM defect in SiC
  with the TM nuclear spin and derive the effective hyperfine tensor within the Kramers doublets formed by the spin-orbit coupling.
  Based on our theory we  discuss the possibility to exchange the nuclear and electron states with potential applications for nuclear spin manipulation and long-lived nunclear-spin based quantum memories.
\end{abstract}

\maketitle

For several applications in quantum technology, such as quantum networks, memories, emitters,
and many more~\cite{kimble08,aharonovich16,heshami16,awschalom21},
a quantum system
needs to be coherently controlled and isolated from unwanted noise at the same time.
Hybrid quantum systems \cite{clerk20,smirnov20,burkard20}, consisting of a part that can couple strongly to external fields
as well as a part that is better shielded from its environment are promising platforms to fulfill this requirement.
In these systems one can benefit from short gate times of one quantum system as well as long coherence times of the other.
A much studied system of this type is the nitrogen vacancy center in diamond with its neighboring nuclear spins \cite{he93,gaebel06,childress06,gali08,felton09,maze11,fuchs11,busaite20,hegde20}
(and Refs.~\cite{doherty13,suter17} for reviews).

Transition metal (TM) defects in silicon carbide (SiC) constitute a similar familiy of systems that have the benefit of being based on a well established host material
as well as having accessible transitions in the telecommunication bands~\cite{bosma18,spindlberger19,gilardoni20,wolfowicz20,csore20}.
Recent studies made the first steps towards control of nuclear spins via transition metal defects in SiC \cite{wolfowicz20}.
While these results are highly promising, a complete theoretical framework is still needed.
In this paper, we derive a model of the hyperfine coupling based on the underlying symmetry properties and relevant orbital configuration of the defect in the crystal,
explaining the experimental data and leading to additional insights.
In particular we derive a sensible form of the interaction of the defect nuclear spin with the  spin and orbital angular momentum of the active electron as well as their combined interaction with external fields.

The prime examples for TM defects in SiC are created by neutral vanadium (V) and positively charged molybdenum (Mo) atoms substituting a Si atom in 6H- or 4H-SiC~\cite{kaufmann97,baur97,bosma18,spindlberger19,gilardoni20,wolfowicz20,csore20}.
These defects have one active electron in the atomic  \(D\)-shell and are invariant under the transformations of the \(C_{3v}\) point group imposed by the crystal structure surrounding the defect.
While the interaction with the nuclear spins of neighboring C and Si isotopes with non-zero nuclear spins is possible,
the presence of such non-zero spin isotopes as a nearest neighbour is fairly improbable,
because their natural abundances are about
1\% for \(^{13}\)C (spin \(1/2\)) and 5\% for \(^{29}\)Si (spin \(1/2\)) \cite{meija16}
and the abundance can be further reduced by using isotopically purified SiC \cite{barbouche17,mazzocchi19}.
Here, we therefore concentrate on the interaction with the TM nuclear spin.
The nuclear spin for the most common V isotope is \(I = 7/2\) (\(>99\%\)) and
\(I = 5/2\) for about \(25\%\) of the stable Mo isotopes and $I=0$ for the remaining isotopes of Mo \cite{audi03,meija16}.

In order to model the hyperfine coupling between the electron and nuclear spins in a TM defect in SiC, we start from the full Hamiltonian
\begin{align}
  \label{eq:Hcomplete}
  H = H_{\mathrm{el}}+ H_{\mathrm{hf}} + H_{z, \mathrm{nuc}} + H_{d, \mathrm{nuc}},
\end{align}
where $H_{\mathrm{el}}=H_{\mathrm{TM}} + V_{\mathrm{cr}} + H_{\mathrm{so}} + H_z + V_{\mathrm{el}} + H_d$  describes the electronic orbital and spin degrees of freedom without their interaction with the nuclear spin \cite{tissot21}, while the remaining terms incorporate the nuclear spin and its interaction with the electron and external fields.
\begin{figure}
\centering
\includegraphics[width=8.5cm]{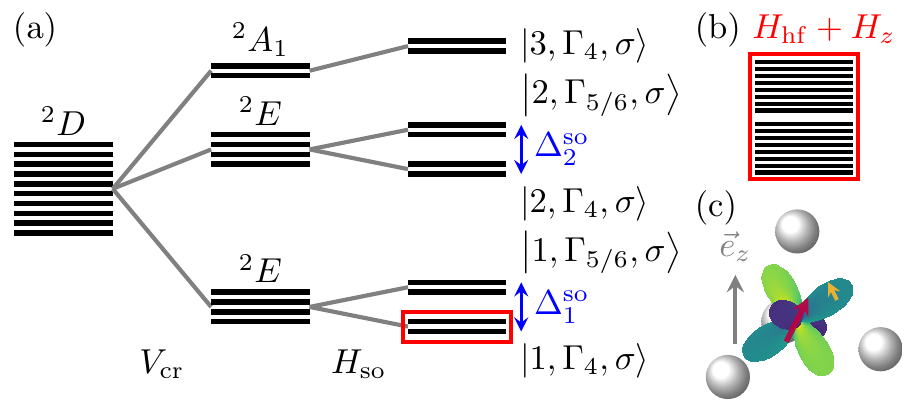}
\caption{\label{fig:Eso} Spin-orbit (a) and hyperfine (b) energy level structure of the active electron bound to the transition metal (TM) defect.
The artistic illustration (c) shows
  the electron with spin-$1/2$ (yellow arrow) occupying a \(D\)-shell (green and violet) which is split by the \(C_{3v}\) symmetric crystal potential,
  arising from the surrounding crystal atoms (white balls),
  into two orbital doublets \(E\) and an orbital singlet \(A_1\) (a).
  The spin-orbit interaction further splits each of the orbital doublets  into two Kramers doublets (KDs),
  leading to the final spin-orbit structure given by five KDs.
 These KDs are then further split into hyperfine levels, due to the interaction with the TM nuclear spin [purple arrow in (c)], shown in (b) for the KD in the red frame.}
\end{figure}
The first three terms in $H_{\mathrm{el}}$ describe the zero-field spin-orbit level structure, as shown in Fig.~\ref{fig:Eso}.
The defect atomic Hamiltonian $H_{\mathrm{TM}}$ localizes the active electron in the \(d\) orbital;
the crystal potential \(V_{\mathrm{cr}}\) reduces the symmetry of the defect to \(C_{3v}\) due to the potentials of the surrounding crystal atoms,
thereby splitting the \(D\)-shell levels;
the spin-orbit Hamiltonian $H_{\mathrm{so}}$ takes the coupling of spin and orbital angular momentum into account and splits the levels further.
In total, this leads to five doubly degenerate levels forming Kramers doublets (KD), pairs of  states related by time-reversal symmetry \cite{dresselhaus10}.
The time-reversal symmetry protects KDs from coupling via operators that are invariant under time-reversal.

The Zeeman Hamiltonian $H_z$ describes the coupling to a static magnetic field which breaks time-reversal symmetry and thus energetically splits the KDs,
the electric potential $V_{\mathrm{el}}$ denotes the coupling to a static external electric field,
and the driving Hamiltonian $H_d$ denotes oscillatory external fields.
Because $V_{\mathrm{el}}$ is invariant under time-inversion, electric fields cannot lift the degeneracy of the KDs,
therefore we concentrate on static magnetic fields in the following.

Typically, the \(C_{3v}\) symmetric crystal potential is sufficiently large to approximate the spin-orbit coupled electronic system by five pseudo-spin subspaces (the KDs) with
distinct zero-field energies $E_{j,\Gamma_{5/6}}, E_{i,\Gamma_4}$ 
where we label the orbital configuration with $j=1,2$ and $i=1,2,3$ and the irreducible representation
(irrep)  of \(C_{3v}\) pertaining to the KD with $\Gamma_{\gamma}$ \cite{tissot21}.
%
The effective Hamiltonian is already diagonal inside blocks of the same orbital configuration for static fields aligned with the crystal axis, i.e. the threefold rotation axis of \(C_{3v}\) which we refer to as the \(z\)-axis in the following.
The reason for this is that a parallel magnetic field only lifts the time-reversal symmetry but keeps the spacial symmetry intact.
On the other hand, fields perpendicular to the crystal axis can mix KDs of the same orbital doublet \(j\) (denoted \(^{2}E\) in Fig.~\ref{fig:Eso}).
The mixing is suppressed by the spin-orbit splitting
\begin{align}
  \label{eq:Dso}
  \Delta^{\mathrm{so}}_j = E_{j, \Gamma_{5/6}} - E_{j, \Gamma_4}
\end{align}
and therefore can be neglected for small fields $\abs{\vec{B}_{\perp}} \ll \min_{j=1,2} \Delta^{\mathrm{so}}_j / \mu_B g_s$.
The coupling between states originating from  different atomic orbitals is even smaller due to the large crystal field splitting \(E_{i,\Gamma_{\gamma}} - E_{i',\Gamma_{\gamma'}}\).

In the following we assume a negligible mixing of the KDs,
as we believe this is the most relevant case for experimental setups
and further technical applications such as quantum memories,
due to the better protection from noise via restricted coupling as a consequence of the intact symmetry.
In this case we can treat each of the KDs as a separate pseudo-spin system, described by the Hamiltonians
acting in the space of the pseudo-spin states $\ket{i, \Gamma_{\gamma}, \uparrow}$ and $\ket{i, \Gamma_{\gamma}, \downarrow}$ of the KD
\begin{align}
  \label{eq:HKDso}
  H^{\mathrm{KD}}_{i, \Gamma_{\gamma}} = E_{i, \Gamma_{\gamma}}  + \frac{\mu_B}{2} \vec{B} \mathbf{g}_{i,\Gamma_{\gamma}} \vec{\sigma}_{i,\Gamma_{\gamma}} ,
\end{align}
with the vector of Pauli operators $\vec{\sigma}_{i,\Gamma_{\gamma}}$,
the Bohr magneton \(\mu_B\), and
the pseudo-spin $g$-tensor $\mathbf{g}$.
For all KDs $\mathbf{g}_{i,\Gamma_{\gamma}}$ is diagonal,
for the $\Gamma_{5/6}$ KDs only $g_{i,\Gamma_{5/6}}^{\parallel} \neq 0$,
for the $\Gamma_4$ KDs with  $j=1,2$ one finds \cite{tissot21}  $g_{j,\Gamma_{4}}^{\parallel} \gg g_{j,\Gamma_{4}}^{\perp}$ if the spin-orbit coupling strength is much smaller than the crystal level spacing.
Using the projection on to the subspace of the KD \(P_{i,\Gamma_{\gamma}}\) we can combine these Hamiltonians to obtain the complete effective spin-orbit Hamiltonian as
\(H_{\mathrm{so}}^{\mathrm{eff}} = \sum_{i, \Gamma_{\gamma}} P_{i,\Gamma_{\gamma}} H^{\mathrm{KD}}_{i, \Gamma_{\gamma}}P_{i,\Gamma_{\gamma}} \).

Now we incorporate the interaction with the nuclear spin given by  Eq.~\eqref{eq:Hcomplete}.
Taking both the Fermi contact and anisotropic hyperfine interaction as well as the orbital nuclear interaction into account \cite{munzarova00,coish09,micera11,vicha14,gohr15},
the total hyperfine Hamiltonian can be written as
\begin{align}
  \label{eq:Hhf}
  H_{\mathrm{hf}} & = H_{\mathrm{FC}} + H_{\mathrm{ahf}} + H_{\mathrm{orb}} \notag \\
  & = \left\{ a_{\mathrm{FC}} \vec{S} + a \left[ \vec{S} - 3 \left( \vec{e}_r \cdot \vec{S} \right) \cdot \vec{e}_r - \vec{L} \right] \right\} \cdot \vec{I},
\end{align}
with the electron (nuclear) spin  $\vec{S}$ ($\vec{I}$) and orbital angular momentum $\vec{L}$ in units of the reduced Planck constant $\hbar$,  the electron direction operator \( \vec{e}_r = \vec{r} / \abs{\vec{r}} \),
and the anisotropic hyperfine and Fermi contact coupling strengths $a = {g_s \mu_{B} \mu_0 g_N \mu_N}/{4\pi r^{3}}$ and $a_{\mathrm{FC}} = - {2 g_s \mu_{B} \mu_0 g_N \mu_N \delta(r)}/{3}$,
with the electron (nuclear) \(g\)-factor $g_s$ ($g_N$), and the nuclear magneton $\mu_N$.
The anisotropic coupling strength depends on the electronic state via $1/r^3$
while the Fermi contact interaction depends on the spin polarization density at the position of the nucleus
denoted using the delta distribution \(\delta(r)\).
The electron is mainly localized in a \(D\)-shell but the mixing with $s$-orbitals
can still lead to a relevant Fermi contact interaction, which is known for TM complexes~\cite{munzarova00,micera11,vicha14,gohr15}.

The nuclear spin can couple to external magnetic fields described by the nuclear Zeeman Hamiltonian
\begin{align}
  \label{eq:Hznuc}
  H_{z,\mathrm{nuc}} = \mu_N g_N \vec{I} \cdot \vec{B} ,
\end{align}
and the corresponding driving \(H_{d, \mathrm{nuc}}\) term for oscillating magnetic fields.
These terms are small in comparison to the KD Zeeman part
\(\abs{\mu_N g_N} \ll g_{i,\Gamma_{\gamma}}^{\parallel} \mu_B\) \cite{baur97,stone05,wolfowicz20}
and diagonal for $\vec{B}$ parallel to the crystal axis.


While the state describing the active electron shows the transformation properties of a \(d\) orbital,
due to effects such as the Jahn-Teller effect and covalency \cite{csore20,ham65,ham68}
there can be an admixture of other orbitals.
The Wigner-Eckart theorem \cite{cornwell97,maze11} enables us to absorb these effects as well as the radial part of the wave-function in reduced matrix elements, in particular to find the minimal set of non-zero matrix elements of (mixed) square components of \(\vec{e}_r\) in Eq.~\eqref{eq:Hhf}, into the orbital basis.
Here, we treat reduced matrix elements as parameters that can be obtained experimentally or via ab-inito calculations.
Then we (perturbativly) transform to the block diagonal basis of the pure spin-orbit Hamiltonian, where spin and orbital states are entangled.
More details are given in the supplemental material~\cite{SM}.

As we did for the mixing due to external fields
we neglect off-diagonal blocks between different orbital configurations \(i\) due to the crystal field splitting.
We find for the effective hyperfine Hamiltonians inside the KDs,
\begin{figure}
  \centering
  \includegraphics[width=8.5cm]{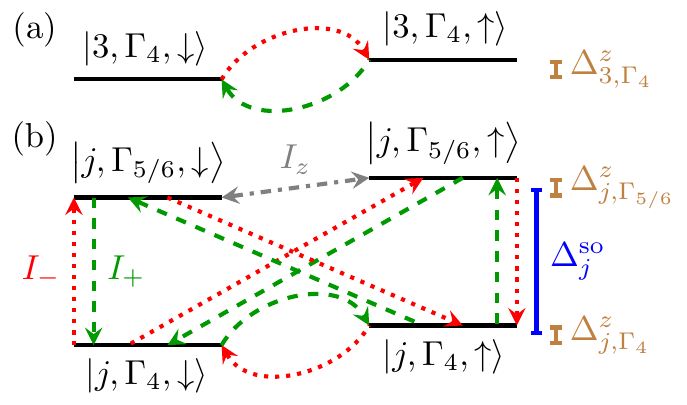}
\caption{ \label{fig:Hhf} Non-zero matrix elements of $H_{\mathrm{hf}}^{\mathrm{eff}}$ between states of two KDs originating from the same orbital doublet.
The matrix elements between states of the same \(\Gamma_4\) KD are proportional to \(I_{\pm}\) (red dotted and green dashed arrows) 
while they are proportional to \(I_z\) (grey dash-dotted arrow) between states of the \(\Gamma_{5/6}\) KDs.
Inside the KDs the (off-diagonal) hyperfine interaction competes with the Zeeman splitting (brown).
The matrix elements between states of the different KDs of the same orbital doublet are proportional to \(I_{\pm}\) and suppressed by the spin-orbit splitting \(\Delta_j^{\mathrm{so}}\) (blue).}
\end{figure}
\begin{align}
  \label{eq:HhfKD1}
   H^{\mathrm{hf}}_{j,\Gamma_{5/6}} & = \frac{1}{2} \left( a^{\parallel}_{j,\Gamma_{5/6}} \sigma_{j,\Gamma_{5/6}}^{z} + a^{\perp}_{j,\Gamma_{5/6}} \sigma_{j,\Gamma_{5/6}}^x \right) I_z , \\
  \label{eq:HhfKD2}
   H^{\mathrm{hf}}_{j,\Gamma_4} & = \frac{a^{\parallel}_{j,\Gamma_4}}{2}  \sigma_{j,\Gamma_4}^z I_z + \frac{a^{\perp}_{j,\Gamma_4}}{4} ( \sigma_{j,\Gamma_4}^+ I_+ + \sigma_{j,\Gamma_4}^- I_- ) , \\
  \label{eq:HhfKD3}
   H^{\mathrm{hf}}_{3,\Gamma_4} & = \frac{a^{\parallel}_{3,\Gamma_4}}{2} \sigma_{3,\Gamma_4}^z I_z + \frac{a^{\perp}_{3,\Gamma_4}}{4} ( \sigma_{3,\Gamma_4}^+ I_- + \sigma_{3,\Gamma_4}^- I_+ ) ,
\end{align}
for \(j=1,2\) and
using the pseudo-spin \(\sigma_{i,\Gamma_{\gamma}}^{\pm} = \sigma_{i,\Gamma_{\gamma}}^x \pm \mathrm{i} \sigma_{i,\Gamma_{\gamma}}^y\)
as well as nuclear \(I_{\pm} = I_x \pm \mathrm{i} I_y\) ladder operators.
The diagonal part of the total effective hyperfine Hamiltonian can thus be written as
\(H_{\mathrm{hf,bd}} = \sum_{i,\Gamma_{\gamma}} P_{i,\Gamma_{\gamma}}H^{\mathrm{hf}}_{i,\Gamma_{\gamma}}P_{i,\Gamma_{\gamma}}\).
The terms mixing the KDs of the same orbital doublet are
\begin{align}
  \label{eq:Hhfod}
  H_{\mathrm{hf,od}} & = \sum_{j,\sigma}
a_{j,c} \Big( \sigma \ket{j, \Gamma_{5/6}, \sigma}\bra{j, \Gamma_4, \sigma} I_{\sigma} \notag \\
& + a_{j,f} \ket{j, \Gamma_{5/6}, \sigma}\bra{j, \Gamma_4, - \sigma} I_{-\sigma} + \mathrm{h.c.} \Big) ,
\end{align}
where $\sigma=\pm 1=\uparrow,\downarrow$.
Combined this leads to the effective hyperfine Hamiltonian \(H_{\mathrm{hf}}^{\mathrm{eff}} = H_{\mathrm{hf,bd}} + H_{\mathrm{hf,od}} \).
The resulting coupling structure is depicted in Fig.~\ref{fig:Hhf}.

Combined with the effective spin-orbit Hamiltonians of the KDs \eqref{eq:HKDso} and nuclear Zeeman Hamiltonian \eqref{eq:Hznuc},
we obtain our first main result,
\begin{align}
  \label{eq:Heff}
  H_{\mathrm{eff}} = H_{\mathrm{so}}^{\mathrm{eff}} + H_{\mathrm{hf}}^{\mathrm{eff}} + H_{z,\mathrm{nuc}}.
\end{align}
The immediate implications of Eq.~\eqref{eq:Heff} are given by its projection on the KDs,
corresponding to the effective description for negligible inter-KD mixing \(\abs{a_{j,c}}, \abs{a_{j,f}} \ll \left|\abs{\Delta^{so}_j} - \sum_{\gamma} \mu_B \abs{g_{i,\Gamma_{\gamma}}^{\parallel} B_{\parallel}} / 2\right|\).
The importance of this is further underlined considering measurements by Wolfowicz \textit{et al.} \cite{wolfowicz20} where the hyperfine coupling strength
is at least two orders of magnitude smaller than the spin-orbit splitting in all V defects in 4H- and 6H-SiC for the ground state ($j=1$).

We now discuss the projection of Eq.~\eqref{eq:Heff} onto the KDs in more detail.
Because the \(\Gamma_4\) KDs transform in complete analogy to pure spin states, the coupling has a familiar form in this case.
In particular the effective hyperfine coupling in the \(\ket{3, \Gamma_4, \sigma}\) KD is the most similar to the simple diagonal dipolar coupling
to the nuclear spin, because the crystal potential does not mix the orbital singlet (\(m=0\)) state with the remaining orbital states
and the spin-orbit coupling vanishes in first order in the orbital singlet.
Additionally, the form of the symmetry allowed part of the anisotropic hyperfine tensor in this case also agrees with that of the \(^{14}\)N NV\(^-\)-center which has the same symmetry but comprises spins \(S=I=1\) \cite{he93,felton09,busaite20}.

The remaining KDs deviate significantly from this form because their two pseudo-spin states have a mixed spin and orbital wavefunction,
due to the interplay of the crystal potential and the spin-orbit coupling.
This leads to the (pseudo-)spin-non-conserving coupling, i.e. the non-diagonal coupling \(\sigma_{j,\Gamma_{5/6}}^{x} I_z\) of the \(\Gamma_{5/6}\) KDs
as well as the \(\sigma_{j, \Gamma_{4}}^{+(-)} I_{+ (-)}\) coupling of the \(\Gamma_4\) KDs for \(j=1,2\), see Fig.~\ref{fig:Hhf}.
Furthermore, the magnitude of \(a_{j, \Gamma_{\gamma}}^{\parallel}\) can deviate significantly from the other two diagonal entries because it can have pure spin contributions.

Group theory further implies that the \(\Gamma_{5/6}\) states cannot be coupled by operators transforming according to the \(E\) representation of \(C_{3v}\), e.g. \(I_x, I_y\).
This follows from the requirement that in $C_{3v}$
the spin-orbit operator part of $H_{\mathrm{hf}}$ has to transform according to the same basis vector of the same irrep as the corresponding nuclear spin operator, because $H_{\mathrm{hf}}$ as a whole has to transform according to $A_1$.
On the other hand the counterpart of \(I_z\) (transforming according to \(A_2\))
can couple these states.
Finally, we stress that the pseudo-spin matrices are not angular momentum type operators and, therefore, \(\sigma_{j,\Gamma_{5/6}}^{x}\) and \(\sigma_{j,\Gamma_{5/6}}^{z}\)
can in part transform according to \(A_2\).
For \(\sigma_{j,\Gamma_{5/6}}^{x}\) the relevant terms in the hyperfine Hamiltonian are \(xy S_y - (y^2-x^2) S_x / 2 \).
We calculate the second order hyperfine interaction inside the KDs due to the interaction between KDs from the same orbital doublet
with a Schrieffer-Wolff transformation~\cite{bravyi11}.
The unperturbed Hamiltonian \(H_0 = \sum_{i,\Gamma_{\gamma}} E_{i,\Gamma_{\gamma}} P_{i,\Gamma_{\gamma}}\)
becomes perturbed inside the KDs
by \(V_{d} = H_{\mathrm{hf,bd}} + H_{z,\mathrm{nuc}} + \sum_{i,\Gamma_{\gamma}} \mu_B g_{\parallel} B_z \sigma_{i,\Gamma_{\gamma}}^z / 2\)
and between KDs with the same orbital origin by \( H_{\mathrm{hf,od}}\),
leading to the first order of the Schrieffer-Wolff transformation
\begin{align}
  S_1 = \ & \sum_{j,\sigma} \frac{1}{\Delta^{\mathrm{so}}_j}
  \Big( a_{j,c} \sigma \ket{j, \Gamma_{5/6}, \sigma}\bra{j, \Gamma_4, \sigma} I_{\sigma}  \notag \\
  \label{eq:S1hf}
&  + a_{j,f} \ket{j, \Gamma_{5/6}, \sigma}\bra{j, \Gamma_4, - \sigma} I_{-\sigma} - \mathrm{h.c.} \Big),
\end{align}
and the new effective Hamiltonian
\( \tilde{H} = H_0 + V_{\mathrm{d}} + \comm{S_1}{H_{\mathrm{hf,od}}}/2 \).
We obtain the second-order correction
  \( {H^{\mathrm{hf},(2)}_{i,\Gamma_{\gamma}}} = P_{i, \Gamma_{\gamma}} \comm{S_1}{H_{\mathrm{hf,od}}} P_{i, \Gamma_{\gamma}}/2\)
with
\begin{align}
  \label{eq:HhfO2}
{H^{\mathrm{hf},(2)}_{j,\Gamma_{5/6}}} &= -{H^{\mathrm{hf},(2)}_{j,\Gamma_{4}}} 
= a^{\mathrm{od}}_j (I^2 - I_z^2 - I_z), 
\end{align}
for $j=1,2$ and \({H^{\mathrm{hf},(2)}_{3, \Gamma_4}} = 0\),
with the defect-configuration dependent constant \(a^{\mathrm{od}}_j = ({a_{j,c}^2 + a_{j,f}^2})/{\Delta^{\mathrm{so}}_j}\).
Combined with Eqs.~\eqref{eq:HhfKD1}--\eqref{eq:HhfKD3} this leads to the second order of the effective hyperfine Hamiltonians for the KDs
\begin{align}
  \label{eq:Hpsnuc}
  H_{i,\Gamma_{\gamma}} = H^{\mathrm{KD}}_{i,\Gamma_{\gamma}} + H^{\mathrm{hf}}_{i,\Gamma_{\gamma}} + H_{z,\mathrm{nuc}} + {H^{\mathrm{hf},(2)}_{i,\Gamma_{\gamma}}}.
\end{align}
We stress that the second order contribution of the hyperfine interaction is purely diagonal in the basis where $z$ points along the crystal axis.

\begin{figure}
\includegraphics[width=9cm]{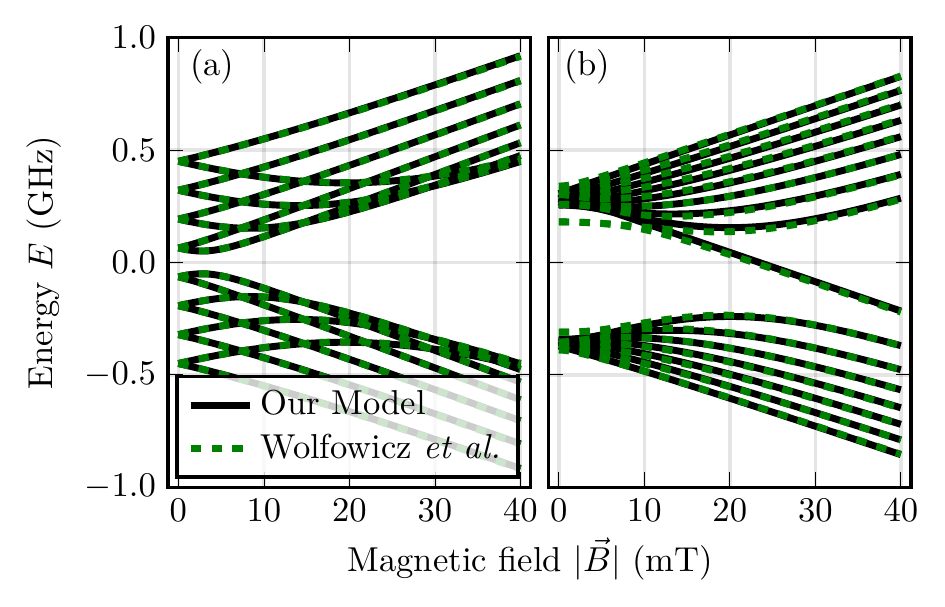}
\caption{\label{fig:EhfKD4}Hyperfine energy levels without zero-field spin-orbit energies.
  For a V defect in the \(\beta\) configuration of \(4H\)-SiC we compare our model (black solid lines), see Eq.~\eqref{eq:Hpsnuc},
  with the fitted model by Wolfowicz \textit{et al.} \cite{wolfowicz20} (green dashed lines).
  The \(\Gamma_{5/6}\) fit values (a) of \cite{wolfowicz20} are completely compatible with our model
  and the energy levels for the \(\Gamma_4\) states (b) were calculated using a least squares fit for the eigenvalues of the models
  for magnetic fields between \(2-45\,\mathrm{mT}\) with 200 data points.
  While there is a disagreement in the energy levels, the allowed transition rates are compatible with the experimental fit.
  The magnetic field of \(2\,\mathrm{mT}\) correponds to the lowest magnetic field visible in \cite{wolfowicz20}.
  For the effective hyperfine constants we find
  \(a_{1,\Gamma_{5/6}}^{\parallel}/h \approx 202.5\, \)MHz and
   \(a_{1,\Gamma_{5/6}}^{\perp}/h \approx 158.2\, \)MHz, as well as
  \(a_{1,\Gamma_4}^{\parallel}/h \approx -174.7 \pm 4.3\, \)MHz and
  \(a_{1,\Gamma_4}^{\perp}/h \approx 149.5 \pm 4.2\, \)MHz, where the fit errors are due to the deviation at small magnetic fields (irrelevant for the transitions).
  We additionally use $g_{1,\Gamma_4,\parallel}=1.87$, $g_{1,\Gamma_{5/6},\parallel}=2.035$, and $\mu_N g_N / h = -11.213\,$MHz/T from \cite{wolfowicz20}.
}
\end{figure}
We now compare our results to the recent measurements by Wolfowicz~\textit{et~al.}~\cite{wolfowicz20}, concentrating on the two ground state KDs.
In \cite{wolfowicz20}, a model for the KDs with a different form of the hyperfine coupling
\(\sum_{k=x,y,z} a_{j,\Gamma_{\gamma}}^k \sigma_{j,\Gamma_{\gamma}}^k I_k / 2 \) was used
for all KDs, additionally allowing a tilt of the quantization axis of the pseudo-spin.
The above-mentioned measurement in combination with our theoretical model suggests that the lowest-energy ground states (GS1) correspond to \(|1, \Gamma_4\rangle\)
and GS2 to \(|1, \Gamma_{5/6}\rangle\).
The first point we want to highlight is that the measurement confirms that the \(|1,\Gamma_{5/6}\rangle\) KD states do not couple via \(I_x,I_y\) and that a tilt of the pseudo-spin around the \(y\)-axis corresponding to the coupling of \(I_z\) to \(\sigma_{j,\Gamma_{5/6}}^x\) is found.
We  highlight that this artificial tilt of the quantization axis needs no further explanation in our theory
where the \( \sigma_{j,\Gamma_{5/6}}^x I_z \) coupling emerges naturally from the interplay of the crystal potential and the spin-orbit interaction.
This can be seen via the transformation properties of the pseudo-spin operators as well as the form of the wavefunction of the KD states.

The second point is that our model provides a resort to explain the measurements \cite{wolfowicz20} for GS1 without the need for an anisotropy in the hyperfine coupling tensor in the plane perpendicular to the crystal axis.
Our model  Eq.~\eqref{eq:HhfKD2}, 
shows good agreement with the transition frequencies in \cite{wolfowicz20},
despite a deviation of two energies in some configurations for small magnetic fields. 
For larger magnetic fields (\(B \ge 10\, \mathrm{mT}\)) indeed all energies are in agreement with the model in \cite{wolfowicz20}.
Not only does our model provide an explanation without the anisotropy, it furthermore reduces the number of free parameters.
For the \(\beta\) configuration of the V defect in \(4H\)-SiC we plot the comparison of the models in Fig.~\ref{fig:EhfKD4}.

Additionally, the measurement of $\beta$ 6H-SiC in \cite{wolfowicz20}
includes all relevant electronic energy splittings
allowing us to assign $\ket{3,\Gamma_4}$ to the lowest energy excited state,
for which we find that the form of the hyperfine interaction of our theory
agrees with their measurement.


Finally we want to use the gained understanding of the effective hyperfine Hamiltonians within the KDs \eqref{eq:Hpsnuc}
and investigate the consequences.
The effective Hamiltonians can be block diagonalised in \(2\times2\) blocks.
In \(\Gamma_{5/6}\) the KDs are mixed with each other but not with the nuclear spin,
such that the resulting states are merely tilted around an axis perpendicular to the crystal axis.
On the other hand, the \(\Gamma_4\) KD electronic states are entangled with the nuclear spin, i.e.
\begin{align}
  \label{eq:mixstate}
 &  \ket{i,\Gamma_4,+,m_I} = \cos(\phi_{i,\Gamma_4,m_I}) \ket{i,\Gamma_4,\uparrow} \ket{m_I}  \\
  &\quad + \sin(\phi_{i,\Gamma_4,m_I}) \ket{i,\Gamma_4,\downarrow} \begin{cases} \ket{m_I-1} \text{ for } i = 1,2 \\ \ket{m_I+1} \text{ for } i=3 \end{cases},\notag
\end{align}
and similarly for the corresponding orthogonal states \(\ket{i,\Gamma_4,-,m_I \mp 1}\).
The analytic diagonalization of the effective hyperfine Hamiltonian~\eqref{eq:Hpsnuc} including a static external magnetic field can be found in the supplemental material \cite{SM}.
\begin{figure}
\centering
\includegraphics[width=8cm]{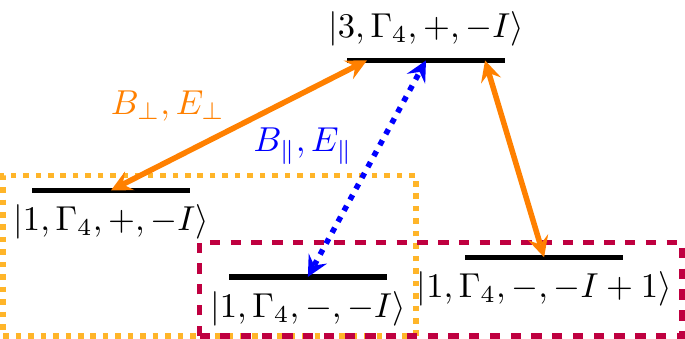}
\caption{\label{fig:Lambda} Lambda (\(\Lambda\)) system to interface the electronic (pseudo-) spin and nuclear spin states (orange arrows).
  Using the employed theory to find the hyperfine eigenstates for a static magnetic field parallel to the crystal axis as well as the selection rules for the spin-orbit eigenstates we obtain the relevant allowed transitions between states of different (pseudo-spin) KDs.
The Lambda system can be used to transfer an electronic state $\alpha \ket{j,\Gamma_4,-,-I} + \beta \ket{j,\Gamma_4,+,-I}$ (yellow dotted frame) to a nuclear spin state (quantum memory, purple dashed frame) $\alpha \ket{j,\Gamma_4,-,-I} + \beta \ket{j,\Gamma_4,-,-I+1}$.
  The blue dotted (solid orange) line(s) corresponds to driving with a magnetic or electric field  parallel (perpendicular) to the crystal axis.}
\end{figure}
In combination with the selection rules for the electronic states \cite{tissot21}, the mixing leads to the relevant allowed transitions.
Here we concentrate on a set of transitions that can be used to transfer the pseudo-spin state of the ground-state KD \(\ket{1, \Gamma_4, \sigma}\) to the nuclear spin and vice versa via a Lambda (\(\Lambda\)) system,
i.e. \(\alpha \ket{1,\Gamma_4,+,-I}+\beta \ket{1,\Gamma_4,-,-I} \leftrightarrow \alpha \ket{1,\Gamma_4,-,-I+1}+\beta \ket{1,\Gamma_4,-,-I}\).
When constructing a \(\Lambda\) system using a different KD
we can neglect the nuclear driving term \(H_{d, \mathrm{nuc}}\) because transitions between the nuclear levels of the same KD pseudo-spin state are highly detuned.
Because \(\ket{3, \Gamma_4, +, -I} = \cos(\phi_{3,\Gamma_4,-I}) \ket{3, \Gamma_4, \uparrow}\ket{-I} + \sin(\phi_{3,\Gamma_4,-I}) \ket{3, \Gamma_4, \downarrow}\ket{-I+1}\),
we immediately see that this state can couple to
\(\ket{1, \Gamma_4, -, -I+1} = \cos(\phi_{1,\Gamma_4,-I}) \ket{1, \Gamma_4, \downarrow} \ket{-I+1} - \sin(\phi_{1,\Gamma_4,-I}) \ket{1, \Gamma_4, \uparrow} \ket{-I}\)
as well as \(\ket{1, \Gamma_4, +, -I} = \ket{1, \Gamma_4, \uparrow} \ket{-I}\)
via a KD pseudo-spin conserving transition.
Therefore, these levels can be used as a \(\Lambda\) system driven by a magnetic or electric field perpendicular to the crystal axis,
as shown in Fig.~\ref{fig:Lambda}.

Analogously our effective theory shows that the hyperfine interaction opens the possibility to directly drive the pseudo-spin transition of the KDs for small magnetic fields due to the pseudo spin tilt or the pseudo-spin nuclear entanglement; this was studied using a different framework by Gilardoni \textit{et al.} \cite{gilardoni21}.
Lastly, when the spin-orbit splitting is sufficiently smal, the second order hyperfine interaction can enable optical driving inside the KDs  by mixing the KDs of the same orbital doublet.

In summary, we introduced a theory to describe the hyperfine interacion in TM defects in SiC having a single electron in a \(D\)-shell.
The theory yields new insights on previous measurements and reduces the required number of fit parameters of the effective hyperfine coupling tensor.
The newly gained insights can be used to construct a \(\Lambda\) transition that can be exploited to create a nuclear spin quantum memory.

\begin{acknowledgments}
  We thank A. Cs{\'o}r{\'e} and A. Gali for useful discussions and
  acknowledge funding from the European Union’s Horizon 2020 research and innovation programme under grant agreement No 862721 (QuanTELCO).
\end{acknowledgments}


\bibliography{refs.bib}


\end{document}